\documentclass[amsmath,amssymb,floatfix, aps, pra, showkeys,twocolumn, superscriptaddress]{revtex4-1}

\usepackage{graphicx}% Include figure files
\usepackage{dcolumn}% Align table columns on decimal point
\usepackage{bm}% bold math
\usepackage{color}
\usepackage{natbib}
\begin{document}

\title{Lost photon enhances superresolution}
\author{A.~Mikhalychev}
\email{mikhalychev@gmail.com}
\affiliation{B.I.Stepanov Institute of Physics, NAS of Belarus, Nezavisimosti ave. 68, 220072 Minsk, Belarus}
\author{P.~Novik}
\affiliation{B.I.Stepanov Institute of Physics, NAS of Belarus, Nezavisimosti ave. 68, 220072 Minsk, Belarus}
\author{I.~Karuseichyk}
\affiliation{B.I.Stepanov Institute of Physics, NAS of Belarus, Nezavisimosti ave. 68, 220072 Minsk, Belarus}
\author{D.~A.~Lyakhov}
\affiliation{Computer, Electrical and Mathematical Science and Engineering Division, 4700 King Abdullah University of Science and Technology, Thuwal 23955-6900, Kingdom of Saudi Arabia}
\author{D.~L.~Michels}
\affiliation{Computer, Electrical and Mathematical Science and Engineering Division, 4700 King Abdullah University of Science and Technology, Thuwal 23955-6900, Kingdom of Saudi Arabia}
\author{D.~Mogilevtsev}
\affiliation{B.I.Stepanov Institute of Physics, NAS of Belarus, Nezavisimosti ave. 68, 220072 Minsk, Belarus}

\date{\today}

\begin{abstract}
Quantum imaging can beat classical resolution limits, imposed by diffraction of light. In particular, it is known that one can reduce the image blurring and increase the achievable resolution by illuminating an object by entangled light and measuring coincidences of photons. If an $n$-photon entangled state is used and the $n$th-order correlation function is measured, the point-spread function (PSF) effectively becomes $\sqrt n$ times narrower relatively to classical coherent imaging. Quite surprisingly, measuring $n$-photon correlations is not the best choice if an $n$-photon entangled state is available. We show that for measuring $(n-1)$-photon coincidences (thus, ignoring one of the available photons), PSF can be made even narrower. This observation paves a way for a strong conditional resolution enhancement by registering one of the photons outside the imaging area. We analyze the conditions necessary for the resolution increase and propose a practical scheme, suitable for observation and exploitation of the effect.
\end{abstract}

\maketitle

%\section{Introduction}

Diffraction of light limits the spatial resolution of classical optical microscopes \cite{abbe1873,rayleigh1879} and hinders their applicability to life sciences at very small scales. Quite recently, a number of superresolving techniques, suitable for overcoming the classical limit, have been proposed. The approaches include, for example, stimulated-emission depletion microscopy \cite{hell1994STED}, superresolving imaging based on fluctuations \cite{dertinger2009fast} or antibunched light emission of fluorescence markers \cite{schwartz2013}, structured illumination microscopy \cite{classen2017superresolution,classen2018analysis}, and quantum imaging \cite{shih_2018_introduction,giovannetti2009sub,xu_experimental_2015}.

Quantum entanglement is known to be a powerful tool for resolution and visibility enhancement in quantum imaging and metrology \cite{shih_2018_introduction,giovannetti2009sub,xu_experimental_2015,boto_quantum_2000,rozema_scalable_2014,giovannetti_quantum-enhanced_2004,agafonov2009high,chan_high-order_2009,chen_arbitrary-order_2010}. It has been shown that using $n$ entangled photons and measuring the $n$-th order correlations, one can effectively reduce the width  of the point-spread function (PSF) $\sqrt{n}$ times \cite{giovannetti2009sub, tsang2009quantum,rozema_scalable_2014,giovannetti_quantum-enhanced_2004} and beat the classical diffraction limit. The increase of the effect with the growth of $n$ can naively be explained as summing up the ``pieces of information'' carried by each photon when measuring their correlations. Such logic suggests that, being given an $n$-photon entangled state, the intuitively most winning measurement strategy is to maximally exploit quantumness of the illuminating field and to measure the maximal available order of the photon correlations (i.e. the $n$th one).

Surprisingly, it is not always the case. First, it is worth mentioning that  effective narrowing of the PSF and resolution enhancement can be achieved  with classically correlated photons \cite{giovannetti2009sub} or even in complete absence of correlations between fields emitted by different parts of the imaged object (as it is for the stochastic optical microscopy \cite{dertinger2009fast}). Moreover, the maximal order of correlations is not necessarily the best one \cite{pearce_precision_2015,vlasenko2020}. In this contribution, we show that, for an entangled $n$-photon illuminating state, it is possible to surpass the measurement of all $n$-photon correlations by loosing a photon and measuring only $n-1$ remaining photons. According to our results, measurement of $(n-1)$th-order correlations effectively leads to $\sqrt{2(n-1)/n}$ times narrower PSF than for commonly considered $n$-photon detection. It is even more strange in view of the notorious entanglement fragility \cite{RevModPhys.81.865}: if even just one of the entangled photons is lost, the correlations tend to become classical. 

The insight for understanding that seeming paradox can be gained from a well-established ghost-imaging technique \cite{pittman_optical_1995,strekalov1995observation,agafonov2009high,chan_high-order_2009,chen_arbitrary-order_2010,bai_ghost_2007,erkmen2008unified,moreau_resolution_2018} and from a more complicated approach of quantum imaging with undetected photons \cite{skornia2001nonclassical,thiel2007quantum,oppel_superresolving_2012,bhatti2018generation}. In our case, detecting $n-1$ photons and ignoring the remaining $n$th one effectively comprises two possibilities (see the imaging scheme depicted in Fig.~\ref{fig:scheme}): the $n$th photon can either fly relatively close to the optical axis of the imaging system towards the detector or go far from the optical axis and fail to pass through the aperture of the imaging system. In the first case, the photon can be successfully detected and provide us its piece of information. In the second case, it does not bring us the information itself, but effectively modifies the state of the remaining $n-1$ photons (as in Refs. \cite{skornia2001nonclassical,thiel2007quantum,bhatti2018generation,cabrillo1999creation,brainis2011quantum}). It effectively produces position-dependent phase shift, thus performing wave-function shaping \cite{brainis2011quantum} and leading to an effect similar to structured illumination \cite{classen2017superresolution,classen2018analysis}, PSF shaping \cite{paur2018tempering}, or linear interferometry measurement \cite{lupo2020quantum}, and enhancing the resolution. We show that for $n > 2$ photons, the sensitivity-enhancement effect leads to higher information gain than just detection of the $n$th photon, and measurement of $(n-1)$-photon correlations surpasses $n$-photon detection.

The discussed sensitivity-enhancement effect can be used for increase of resolution in practical imaging schemes. One can devise a conditional measurement set-up by placing a bucket detector outside the normal pathway of the optical beam (e.g. near the lens outside of its aperture) and post-selecting the outcomes when one photon gets to the bucket detector and the remaining $n-1$ ones successfully reach the position-sensitive detector, used for the coincidence measurements. We show that such post-selection scheme indeed leads to additional increase of resolution relatively to $(n-1)$-photon detection. Resolution enhancement by post-selecting the more informative field configuration is closely related to the spatial mode demultiplexing technique \cite{tsang2016quantum,tsang2017subdiffraction}. However, in our case the selection of more informative field part is performed by detection of a photon while all the remaining photons are detected in the usual way rather than by filtering the beam itself. Also, our technique bears some resemblance to the multi-photon ghost-imaging \cite{agafonov2009high}.

For drawing quantitative conclusions about the resolution enhancement of the proposed technique relatively to traditional measurements of $n$ and $(n-1)$-photon coincidences, we employ the Fisher information, which has already proved itself as a powerful tool for analysis of quantum imaging problems and for meaningful quantification of resolution \cite{motka2016optical,tsang2016quantum,tsang2017subdiffraction,mikhalychev_efficiently_2019,paur2018tempering,paur2019reading,pearce_precision_2015,vlasenko2020}. Our simulations show that for imaging a set of semi-transparent slits (i.e., for multi-parametric estimation problem), one indeed has a considerable increase in the information per measurement, and the corresponding resolution enhancement. While genuine demonstration of the discussed effects requires at least 3 entangled photons, which can be generated by a setup with complex nonlinear processes (e.g. cascaded spontaneous parametric down-conversion (SPDC) \cite{hubel2010direct}, combination of SPDC with up-conversion \cite{keller1998theory}, cascaded four-wave mixing \cite{wen2010tripartite}, or the third-order SPDC \cite{corona2011experimental,corona2011third,borshchevskaya2015three}), a relatively simple biphoton case is still suitable for observing resolution enhancement for a specific choice of the region where the $n$th photon (here, the second one) is detected. 

%The paper is organized as follows. First, we discuss the basic imaging scheme and demonstrate the advantage of measuring $(n-1)$-photon coincidences in terms of PSF narrowing. Then, the resolution enhancement relatively to $n$-photon detection is interpreted in terms of the effective modification of the $n-1$ photons' state according to the ``fate'' of the $n$th photon. The obtained general expressions and the idea of additional resolution enhancement by intentional detection of a photon outside the imaging system are illustrated by a simple model example with just 2 pinholes. Finally, the approach is applied to a realistic problem of imaging a set of semitransparent slits and quantitative analysis of the resolution by means of Fisher information is presented and discussed.

\section{Results}

\subsection{Imaging with entangled photons}

\begin{figure}[htbp]
\includegraphics[width=0.9\linewidth]{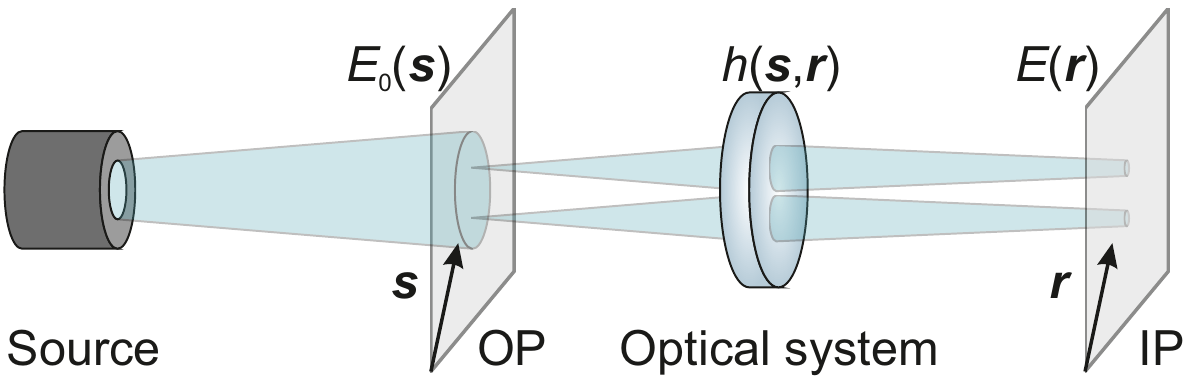}
\caption{General scheme of an imaging setup. See the text for details. An image of an object placed at the object plane OP is formed by the optical system at the image plane IP.}
\label{fig:scheme}
\end{figure}

We consider the following common model of a quantum imaging setup (Fig.~\ref{fig:scheme}). An object is described by a transmission amplitude $A(\vec s)$, where $\vec s$ is the vector of transverse position in the object plane. It is illuminated by linearly polarized light in an $n$-photon entangled quantum state 
\begin{multline}
\label{eq:Psi_n}
|\Psi_n\rangle \propto \int d^2 \vec k_1 \cdots d^2 \vec k_n \delta^{(2)}(\vec k_1 + \cdots + \vec k_n) \\ {} \times a^+ (\vec k_1) \cdots a^+ (\vec k_n) |0 \rangle  \propto \int d^2 \vec s \left(a^+ (\vec s)\right) ^ n |0 \rangle,
\end{multline}
where $a^+ (\vec k)$ and $a^+ (\vec s)$ are the operators of photon creation in the mode with the wavevector $\vec k$ and at position $\vec s$ respectively. An optical system with the PSF $h(\vec s, \vec r)$ maps the object onto the image plane, where the field correlations are detected. 

Features of the field passing through the analyzed object (and, thus, the object parameters) can be inferred from the measurement of intensity correlation functions accomplished by simple coincidence photo-counting. The detection rate of the $n$-photon coincidence at a point $\vec r$ is determined by the value of the $n$th-order correlation function (see Methods for details):
\begin{equation}
\label{eq:Gn}
G^{(n)}(\vec r) \propto \left| \int d^2 \vec s A^n (\vec s) h^n(\vec s, \vec r) \right|^2.
\end{equation}
The signal, described by Eq.~(\ref{eq:Gn}), includes the $n$th power of the PSF, which is $\sqrt n$ times narrower than the PSF itself. At least for the object of just two transparent point-like pinholes, such narrowing 
yields $\sqrt n$ times better visual resolution of the object than for imaging with coherent light (see e.g. Refs. \cite{giovannetti_quantum-enhanced_2004,shih_2018_introduction}). 

Alternatively, one may try to ignore one of the photons and measure correlations of the remaining $(n-1)$ ones. The rate of $(n-1)$-photon coincidences is described by the $(n-1)$th-order correlation function:
\begin{multline}
\label{eq:Gn-1}
G^{(n-1)}(\vec r) \propto  \int d^2 \vec s \left|A (\vec s)\right|^{2(n-1)} \left|h(\vec s, \vec r)\right|^{2(n-1)}.
\end{multline}
Here, the $2(n-1)$th power of the PSF is present. For $n>2$, the resolution enhancement factor $\sqrt{2(n-1)}$ is larger than the factor $\sqrt{n}$ achievable for $n$-photon detection.

The result obtained looks quite counter-intuitive: each photon carries some information about the illuminated object, while discarding one of the photons leads to additional information gain. This seeming contradiction is just a consequence of applying classical intuition to quantum dynamics of an entangled system. Due to quantum correlations, an entangled photon can affect the state of the remaining ones and increase their ``informativity'' even when it is lost without being detected \cite{skornia2001nonclassical,thiel2007quantum,oppel_superresolving_2012,bhatti2018generation}. 
Here we show that in our imaging scheme such an enhancement by loss is indeed taking place.  Moreover, an additional resolution increase can be achieved through conditioning by detecting the photon outside the aperture of the imaging system (see Fig.~\ref{fig:scheme}).

\subsection{Effective state modification}

Let us consider the change of the $(n-1)$ photons state depending on the ``fate'' of the $n$th photon in more details. We follow the approach discussed in Ref. \cite{bhatti2018generation}, which consists in splitting the description of an $n$-photon detection process into 1-photon detection, density operator modification, and subsequent $(n-1)$-photon detection for the modified density operator. If we detect $(n-1)$-photon coincidence, the $n$th photon can be: (1) transmitted through the object into the imaging system aperture, (2) transmitted through the object outside the imaging system aperture, or (3) absorbed by the imaged object. 

For the first possibility, the $n$th photon can reach the detector and potentially be registered at certain point $\vec r'$. The effective state of the remaining photons is (see Methods):
\begin{equation}
\label{eq:Psi_n-1(1)}
|\Psi_{n-1}^{(1)}(\vec r')\rangle \propto \int d^2 \vec s A(\vec s) h(\vec s, \vec r') (a^+(\vec s))^{n-1} |0\rangle.
\end{equation}

If we are interested in detection of all the $n$ photons (i.e. postselect the cases when the $n$th photon is successfully detected at the position $\vec r' = \vec r$), the information gain due to the $n$th photon detection results from the factor $h(\vec s, \vec r)$ introduced into the effective state (\ref{eq:Psi_n-1(1)}) of the remaining photons. It forces the photons to pass through the particular part of the object, which is mapped onto the vicinity of the detection point $\vec r$, and effectively reduces the image blurring.

If, according to the second possibility, the $n$th photon goes outside the aperture of the imaging system and has the transverse momentum component $\vec k$, the effective state of the remaining photons is
\begin{equation}
\label{eq:Psi_n-1(2)}
|\Psi_{n-1}^{(2)}(\vec k)\rangle \propto \int d^2 \vec s A(\vec s) e^{i \vec k \cdot \vec s} (a^+(\vec s))^{n-1} |0\rangle.
\end{equation}

An important feature of Eq. (\ref{eq:Psi_n-1(2)}) is the factor $e^{i \vec k \cdot \vec s}$, which effectively introduces the periodic phase modulation of the field, illuminating the object, and leads to a similar effect as intensity modulation for the structured illumination approach \cite{classen2017superresolution,classen2018analysis}.

To take into account possible absorption of the $n$th photon by the imaged object, one can introduce an additional mode and model the object as a beamsplitter (see e.g. Ref. \cite{lemos2014quantum}). Similarly, to the two previously considered cases, the following expression can be derived for the effective $(n-1)$-photon density operator:
\begin{equation}
\label{eq:rho_n-1(3)}
    \rho_{n-1}^{(3)} = \int d^2 \vec s [1 - |A(\vec s)|^2] (a^+(\vec s))^{n-1} |0\rangle \langle 0 |(a(\vec s))^{n-1}.
\end{equation}

By averaging over the three discussed possibilities (see Methods), one can obtain the following expression for the effective state of the remaining $(n-1)$ photons:
\begin{equation}
\label{eq:rho_n-1_final}
    \rho_{n-1} = \int d^2 \vec s (a^+(\vec s))^{n-1} |0\rangle \langle 0 |(a(\vec s))^{n-1},
\end{equation}
which indicates the well-known effect of turning an $n$-photon entangled state into an $(n-1)$-photon classically correlated state when one of the photons is excluded from consideration.

The detailed derivation of the result, while being quite trivial from formal point of view, helps us to get to the following physical conclusions:
\begin{itemize}
    \item The effective state of $n-1$ photons (and the $(n-1)$th order correlation function) is modified, even if the $n$th photon is not detected, and depends on its ``fate'' (the way the photon actually passes). 
    \item The effective state of  $(n-1)$-photons might be changed in a way, which provides the object resolution enhancement. 
    \item When $n$-photon coincidences are measured, the $n$th photon detection actually leads to the postselection due to discard of possibilities leading to the photon loss.
    \item For $n > 2$, the advantage gained from registering more photon coincidences with the $n$th photon detection does not compensate for the information loss caused by discarding outcomes corresponding to the strongly modified $(n-1)$-photon state, which is more sensitive to the object features (see Eqs. (\ref{eq:Gn}) and (\ref{eq:Gn-1})).
\end{itemize}

Further, we discuss how the advantageous outcomes can be postselected, instead of being discarded, for resolution enhancement.

\subsection{Model example}

\begin{figure}[htbp]
\includegraphics[width=\linewidth]{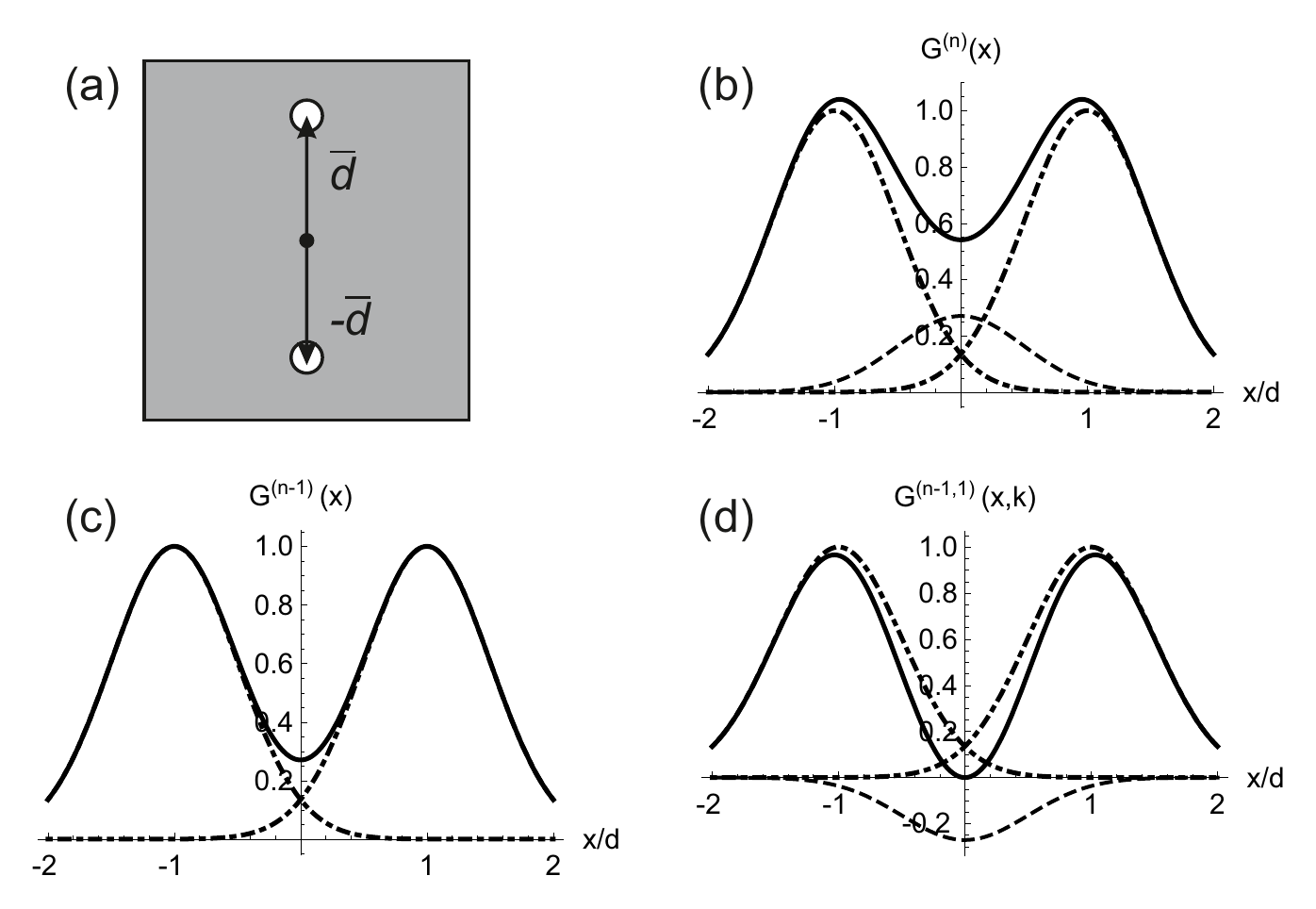}
\caption{Model example: imaging two pinholes (a), separated by the distance $2 d$. Constructive (b), absent (c) and destructive interference (d) of the light passing through the pinholes. Dot-dashed lines represent separate contributions from the pinholes; dashed line shows the interference signal; solid line represents the sum of all the contributions.}
\label{fig:model_example}
\end{figure}

To gain better understanding of the processes of resolution enhancement by a photon loss and postselection, let us consider a standard model object illuminated by an $n$-photon entangled state and consisting of two pinholes, which are separated by the distance $2 d$ and positioned at the points $\vec d$  and $- \vec d$ (Fig.~\ref{fig:model_example}a).  If the pinholes are small enough, the light passing through the object can be decomposed into just two field modes, corresponding to the spherical waves emerging from the two pinholes and further denoted by the indices ``$+$'' and ``$-$'' for the upper and the lower pinhole respectively. For simplicity's sake, we assume that the PSF is a real-valued function,  and that the light state directly after the object has the form of a NOON-state of the discussed modes ``$+$'' and ``$-$'':
\begin{equation}
|\Phi_n \rangle \propto | n \rangle_{+} | 0 \rangle_{-} + | 0 \rangle_{+} | n \rangle _{-}.
\end{equation}

The $n$th-order correlation signal contains separate contributions from the single pinholes and a cross-term, caused by constructive interference and leading to additional blurring of the image (Fig.~\ref{fig:model_example}b):
\begin{equation}
\label{eq:Gn_model}
G^{(n)}(\vec r) \propto h^{2n}(\vec d, \vec r) + h^{2n} (-\vec d, \vec r) + 2 h^n (\vec d, \vec r) h^n (-\vec d, \vec r).
\end{equation}
The $(n-1)$th-order correlations include only separate single-pinhole signals and produce a sharper image (Fig.~\ref{fig:model_example}c):
\begin{equation}
\label{eq:Gn-1_model}
G^{(n-1)}(\vec r) \propto h^{2(n-1)}(\vec d, \vec r) + h^{2(n-1)} (-\vec d, \vec r).
\end{equation}

Let us interpret these results in terms of detecting $n-1$ photons conditioned by the $n$th photon detection. According to Eq. (\ref{eq:Psi_n-1(1)}), if the photon is detected at the point $\vec r$ of the image plane, it transforms the state of the remaining photons into 
\begin{multline}
|\Phi_{n-1}^{(1)}(\vec r) \rangle  \propto h(\vec d, \vec r) | n - 1 \rangle_{+} | 0 \rangle_{-} \\ {} + h(-\vec d, \vec r) | 0 \rangle_{+} | n - 1 \rangle _{-}.
\end{multline}
The state coherence is preserved, while the blurring, caused by constructive interference, is slightly reduced due certain ``which path'' information provided by the $n$th photon detection. 

If the $n$th photon is characterized by the transverse momentum $\vec k$, $|\vec k| > k_{max}$, and does not get into the imaging system aperture, the effective modified state of the remaining photons is (see Eq. (\ref{eq:Psi_n-1(2)})):
\begin{multline}
|\Phi_{n-1}^{(2)}(\vec k) \rangle  \propto e^{i \vec k \cdot \vec d} | n - 1 \rangle_{+} | 0 \rangle_{-} \\ {} + e^{-i \vec k \cdot \vec d} | 0 \rangle_{+} | n - 1 \rangle _{-},
\end{multline}
Now, the phase shift between the two modes depends on $\vec k$ and can lead to destructive interference, which enhances the contrast of the image. For example, when $\vec k \cdot \vec d = \pi / 2$, one has maximally destructive interference and the detected signal is proportional to $|h^{(n-1)}(\vec d, \vec r) - h^{(n-1)} (-\vec d, \vec r)|^2$ with $100\%$ visibility of the gap between the two peaks (Fig.~\ref{fig:model_example}d).

Discarding the information about the $n$th photon (measuring $G^{(n-1)}$) corresponds to averaging over the possibilities to have the photon passing to the detector and missing it, and yields the following mixed state of the remaining photons:
\begin{multline}
\rho_{n-1} \propto |n - 1\rangle_{+} \langle n - 1| \otimes |0\rangle_{-} \langle 0 | \\{} + |0\rangle_{+} \langle 0| \otimes |n - 1\rangle_{-} \langle n - 1 |.
\end{multline}
I.e., the cross-terms with constructive and destructive interference cancel each other, and resulting mixed state allows for some resolution gain over the pure $n$th photon NOON state.

\subsection{Application to quantum imaging}

\begin{figure}[htbp]
\includegraphics[width=\linewidth]{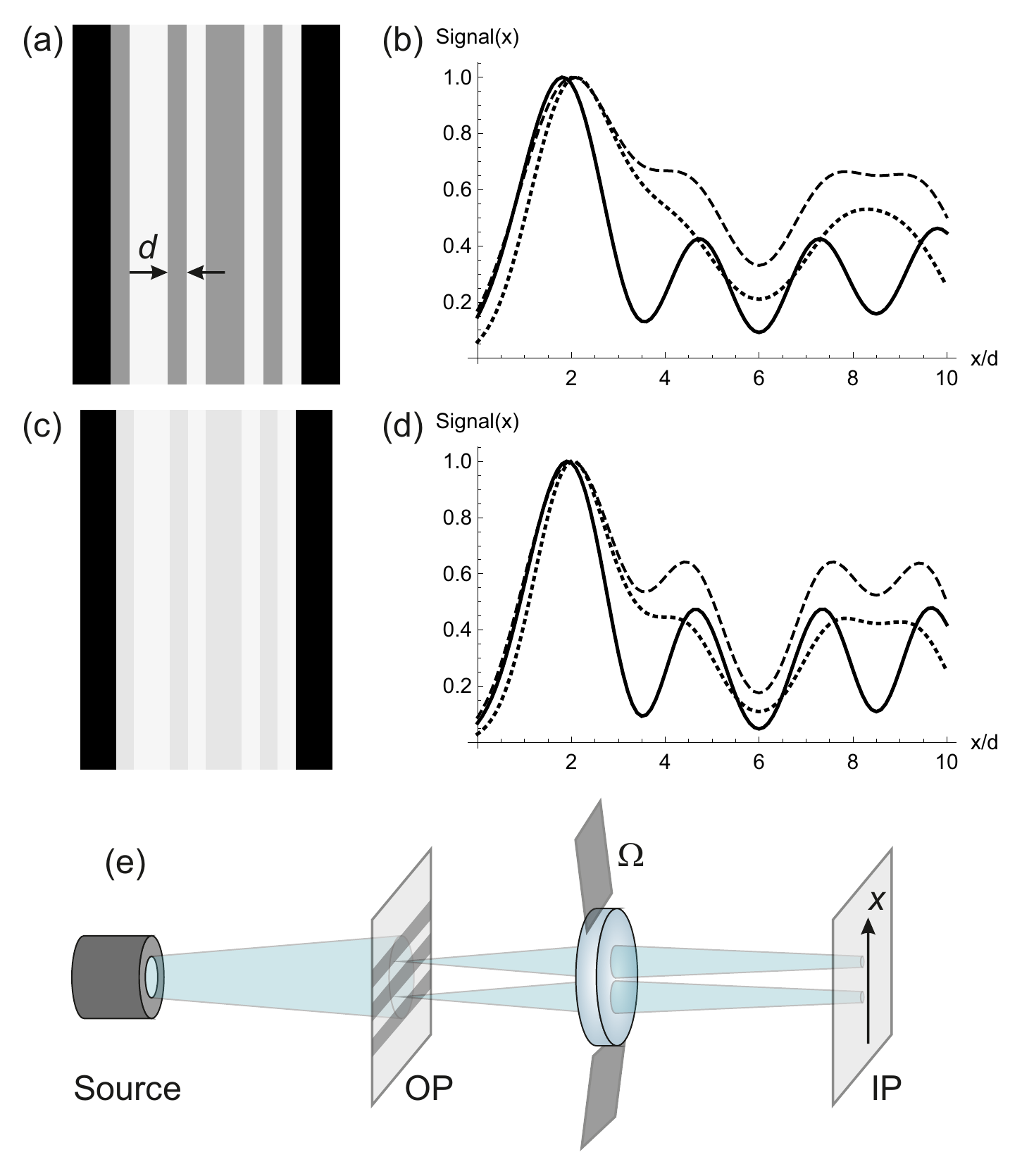}
\caption{Model objects (a, c), imaging scheme for $(n-1)$-photon coincidence conditioned by detecting a photon in the region $\Omega$ (e), and simulation results (b, d). Sets of semitransparent slits with transmission amplitudes $0.5 \div 1$ (a) and $0.9 \div 1$ (c) were used as model objects. The signal ($G^{(n)}$ --- dotted lines, $G^{(n-1)}$ --- dashed lines, $G^{(n-1,1)}$ --- solid lines) was simulated for the object from panel (a) and $n = 3$ (b) and 4 (d). The detection region for the $n$th photon is $\Omega = \{ \vec k \colon k_{max} \le |\vec k| \le 2 k_{max}\}$. The axis $x$ is directed across the slits. The coordinate $x$ is normalized by the slit size $d$ indicated in plot (a).}
\label{fig:simulations}
\end{figure}

To illustrate possible application of the ideas to practical quantum imaging, we consider an object represented by a set of semitransparent slits (Fig.~\ref{fig:simulations}a,c). The resolution of the modeled optical system is limited by diffraction at the lens aperture, which admits only the photons with the transverse momentum $\vec k$ not exceeding $k_{max}$: $|\vec k| \le k_{max}$. 

We compare the following three strategies: (i) measuring $G^{(n)} (\vec r)$ along the direction perpendicular to the slits (the signal is described by Eq.~(\ref{eq:Gn})); (ii) measuring $G^{(n-1)} (\vec r)$ at the same points (Eq.~\ref{eq:Gn-1}); (iii) measuring coincidence signal $G^{(n-1,1)}(\vec r, \Omega)$ of $n-1$ photons detected at the point $\vec r$ of the detection plane and the $n$th photon being anywhere in certain region $\Omega$ outside the lens aperture. For the latter case the signal can be written as
\begin{multline}
G^{(n-1,1)}(\vec r, \Omega) \propto \int d^2 \vec s d^2 \vec s' A^n (\vec s) A^n(\vec s') \\{} \times h^{n-1}(\vec s, \vec r) h^{n-1}(\vec s', \vec r)) g(\vec s - \vec s'),
\end{multline}
where
\begin{equation}
\label{eq:kernel}
g(\vec s - \vec s') = \int_{\vec k \in \Omega} d^2 \vec k e^{i \vec k \cdot (\vec s - \vec s')}.
\end{equation}

Integration in Eq.~(\ref{eq:kernel}) corresponds to detection of the $n$th photon by a bucket detector, similarly to multiphoton ghost imaging \cite{agafonov2009high,chan_high-order_2009,chen_arbitrary-order_2010}. However, all the $n$ photons (including the $n-1$ ones, which get to the position-resolving detector) do pass through the investigated object. Our scheme can also be considered as a generalization of hybrid near- and far-field imaging, when the entangled photons are analyzed partially in position space and partially in momentum space.
%Also, the scheme can be considered as a generalization of hybrid near- and far-field imaging [Ref...], but with $n$ photons and a bulk detector in the far field.

Simulated images are shown in Fig.~\ref{fig:simulations}b,d. One can clearly see that $(n-1)$-photon detection yields better visual resolution than measurement of the $n$th-order coincidences, while $G^{(n-1,1)}$ provides additional contrast enhancement.

Of course, the effective narrowing of the PSF by measuring correlation functions does not necessarily mean a corresponding increase in precision of inferring of the analyzed parameters (positions of the object details, channel characteristics, etc.) \cite{pearce_precision_2015,vlasenko2020}. However, at least for certain imaging tasks (such as, for example, a cornerstone problem of finding a distance between two point sources in the far-field imaging), narrowing of the PSF can indeed lead to increase of the informational content per measurement, and to the potentially unlimited resolution with increasing of $n$ \cite{vlasenko2020}.

To describe the resolution enhancement in a quantitative and more consistent way, we employ Fisher information. Let the transmission amplitude of the object be decomposed as $A(\vec s) = \sum_\mu \theta_\mu f_\mu (\vec s)$, where the basis functions $f_\mu (\vec s)$ can represent e.g. slit-like pixels for the considered example \cite{mikhalychev_efficiently_2019}. Then the problem of finding $A(\vec s)$ becomes equivalent to reconstruction of the unknown decomposition coefficients $\theta_\mu$. If one has certain signal $S(\vec r)$, sampled at the points $\{\vec r_i\}$, Fisher information matrix (FIM) \cite{fisher1925theory,rao1945information}, normalized by a single detection event, can be introduced as
\begin{equation}
\label{eq:Fisher}
F_{\mu\nu} = \sum_i\left(\frac{1}{S(\vec r_i)} \frac{\partial S(\vec r_i)}{\partial \theta_\mu} \frac{\partial S(\vec r_i)}{\partial \theta_\nu} \right) / \sum_i S(\vec r_i).
\end{equation}

Cram{\'e}r-Rao inequality \cite{cramer1946,rao1945information} bounds the total reconstruction error (the sum of variances of the estimators for all the unknowns $\{\theta_\mu\}$) by the trace of the inverse of FIM:
\begin{equation}
\label{eq:Delta2}
\Delta^2 = \sum_\mu \Delta \theta_\mu \ge \frac{1}{N} \operatorname{Tr} F^{-1},
\end{equation}
where $N$ is the number of registered coincidence events. When the size of the analyzed object features (e.g. the slit size $d$ in Fig.~\ref{fig:simulations}a) tends to zero, the bound in Eq.~(\ref{eq:Delta2}) diverges (the effect is termed ``Rayleigh's curse''). The achievable resolution can, therefore, be determined by the feature size $d$, for which $\operatorname{Tr} F^{-1}$ starts growing rapidly with the decrease of $d$. A more rigorous definition can be given by specifying certain reasonable threshold $N_{max}$ for the maximal required number of registered coincidence events $N$ (e.g. we take $N_{max} = 10^5$ for further examples) and imposing the restriction $\operatorname{Tr} F^{-1} \le N_{max}$ (see Methods).

\begin{figure}[htbp]
\includegraphics[width=\linewidth]{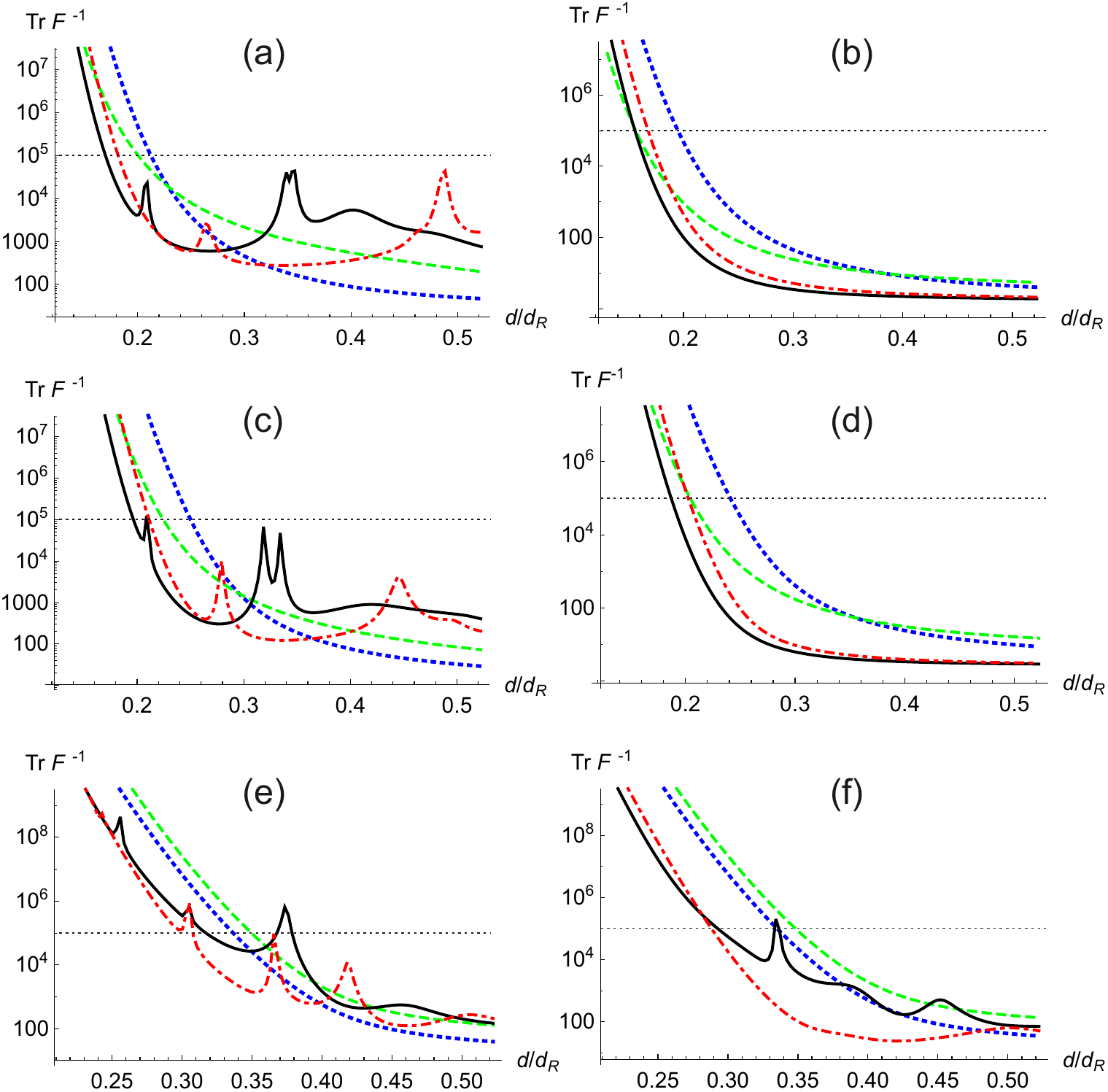}
\caption{Dependence of the trace of inverse FIM on the normalized slit size $d / d_R$ for detection of $n$th-order correlations (blue dotted lines), $(n-1)$th-order correlations (green dashed lines), and $(n-1)$th-order correlations, conditioned by detection of the $n$th photon in the region $\Omega = \{ \vec k \colon k_{max} \le |\vec k| \le 2 k_{max}\}$ (black solid lines) and $\Omega = \{ \vec k \colon k_{max} \le |\vec k| \le 1.5 k_{max}\}$ (red dot-dashed lines) for $n = 4$ (a,b), 3 (c,d), and 2 (e,f). The objects are shown in Fig.~\ref{fig:simulations}a (for plots a, c, and e) and Fig.~\ref{fig:simulations}c (for plots b, d, and f). Horizontal dotted lines indicate the threshold $\operatorname{Tr}F^{-1} \le N_{max} = 10^5$, used for quantification of resolution.}
\label{fig:Fisher}
\end{figure}

The dependence of the predicted reconstruction error on the normalized object scale $d / d_R$ (where $d_R = 3.83 / k_{max}$ is the classical Rayleigh limit for the considered optical system) is shown in Fig.~\ref{fig:Fisher}. The sampling points for the signal are taken with the step $d/2$ along a line perpendicular to slits. As expected, ignoring $n$th photon and measuring $(n-1)$-photon coincidences brings about larger achievable information per measurement, and correspondingly lower errors in object parameter estimation, yielding $(10\div20)\%$ better resolution for $n = 3$ and 4. According to the theoretical predictions, for $n = 2$ no resolution increase is observed. 

Also, our results confirm that the proposed hybrid scheme is indeed capable of increasing the resolution for $n = 3$ and 4 by conditioned detection of the $n$th photon. An additional information that can be gained relatively to the measurement of $(n-1)$-photon correlations is about $(10\div 15)\%$. However, that gain vanishes in the regime of deep superresolution ($d \lesssim 0.2 d_R$): the black solid and red dot-dashed lines intersect with the green dashed one for small $d / d_R$ in Fig.~\ref{fig:Fisher}. The reason for such behavior is that the effective phase shift, introduced by detection of the $n$th photon with the transverse momentum $\vec k$ outside of the aperture of the imaging system, becomes insufficient for the resolution enhancement for $|\vec k| d \ll 1$.

For $n=2$, the scheme also can give certain advantages (Fig. \ref{fig:Fisher}e,f), which, however, are not so prominent because they do not stem from the fundamental requirement of having better resolution for $G^{(n-1)}$ than for $G^{(n)}$. Still, taking into account the difficulties in generation of 3-photon entangled states \cite{hubel2010direct,keller1998theory,wen2010tripartite,corona2011experimental,corona2011third,borshchevskaya2015three}, an experiment with biphotons can be proposed for initial tests of the approach.

\begin{figure}[htbp]
\includegraphics[width=0.6\linewidth]{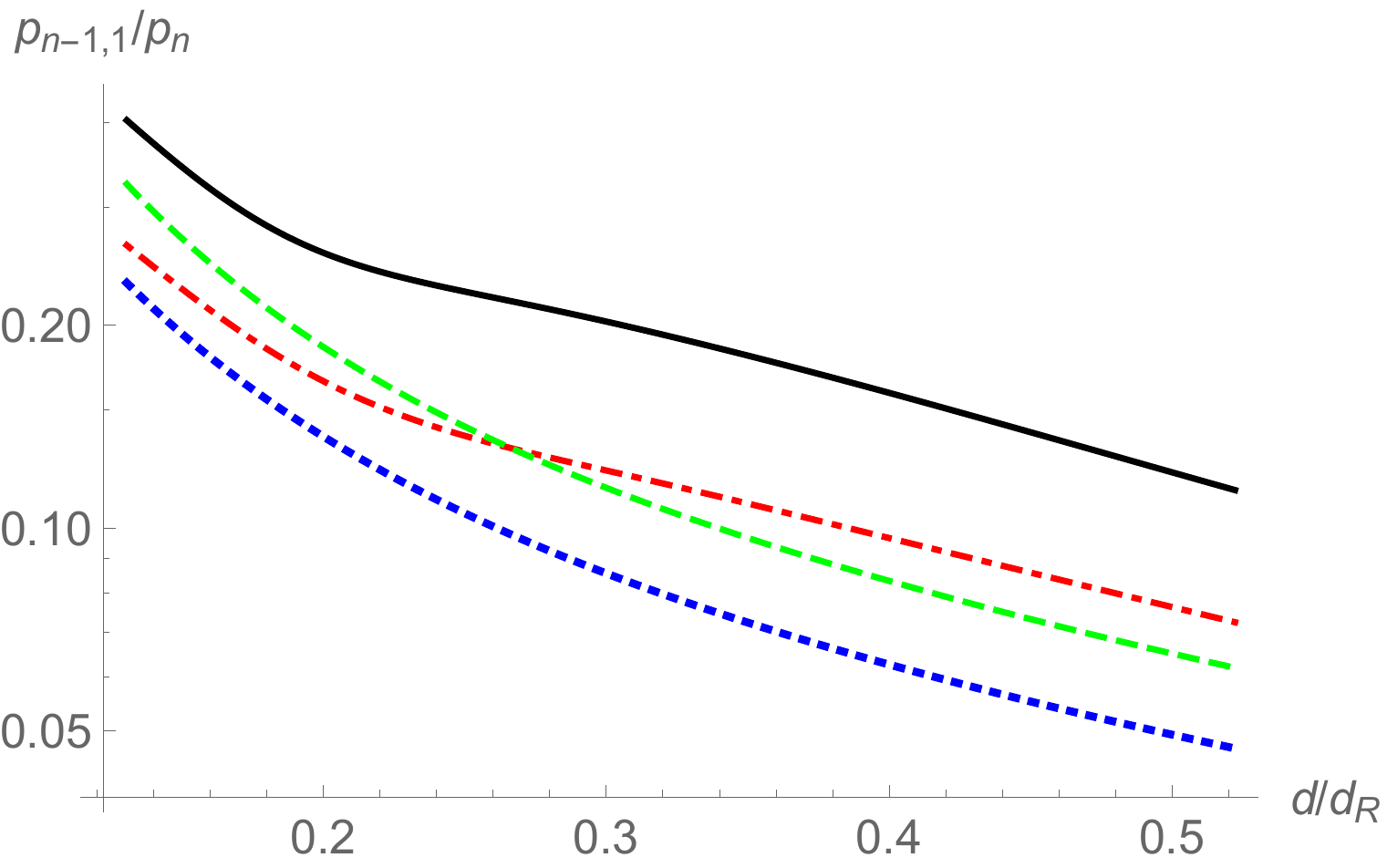}
\caption{Dependence of the overall detection probabilities ratio  on the normalized slit size $d / d_R$ (see details in the text). The $n$th photon is detected in the region $\Omega = \{ \vec k \colon k_{max} \le |\vec k| \le 2 k_{max}\}$ (solid black and dashed green lines) or $\Omega = \{ \vec k \colon k_{max} \le |\vec k| \le 1.5 k_{max}\}$ (dot-dashed red and dotted blue lines) for $n = 4$. The objects are shown in Fig.~\ref{fig:simulations}a (for solid black and dot-dashed red lines) and Fig.~\ref{fig:simulations}c (for dashed green and dotted blue lines). Horizontal dotted lines indicate the threshold, used for quantification of resolution.}
\label{fig:rate}
\end{figure}

The plots, shown in Fig. \ref{fig:Fisher}, represent information per single detection event. Therefore, certain concerns about the rates of such events may arise: waiting for a highly informative, but very rare event can be impractical. Fig. \ref{fig:rate} shows the ratio of the overall detection probabilities $p_{n-1,1} / p_n$, where $p_{n-1,1}$ corresponds to $(n-1)$-photon coincidence, conditions by detection of a photon outside the aperture, and $p_n$ describes traditional measurement of $n$-photon coincidences. For a signal $S(\vec r_i)$, the overall detection probability is defined as $p = \sum_i S(\vec r_i)$ and represents the denominator of Eq. (\ref{eq:Fisher}). When plotting Fig. \ref{fig:rate}, we do not include the measurement of $G^{(n-1)}$ into the comparison, because the ratio of probabilities for $(n-1)$ and $n$-photon detection events  strongly depends on details of a particular experiment, such as the efficiency of the detectors. 

The rate of $(n-1)$-photon coincidences, conditioned by the detection of a photon outside of the aperture, is indeed $3\div 20$ times smaller than the rate of $n$-photon coincidences. However, for the considered multi-parametric problem, the ``Rayleigh curse'' leads to very fast decrease of information when the slit size $d$ becomes smaller than the actual resolution limit, and the effect of rate difference is almost negligible. For example, for $n = 4$, the object, shown in Fig. \ref{fig:simulations}a, and the threshold $\operatorname{Tr}F^{-1} \le 10^5$, the minimal slit width $d$ for successful resolution of transmittances equals $0.212 d_R$ for the measurement of $G^{(n)}$, $0.170 d_R$ for the measurement of $G^{(n-1,1)}$ with the $n$th photon detected in the region $\Omega = \{ \vec k \colon k_{max} \le |\vec k| \le 2 k_{max}\}$, and $0.177 d_R$ for the same measurement of $G^{(n-1,1)}$ when the reduced detection rate is taken into account. Notice, that all the mentioned values of the resolved feature size $d$ are quite far beyond the classical resolution limit $d_R$.

At the first glance, the reported percentage of the resolution enhancement does not look very impressive or encouraging. However, one should keep in mind that the increase of the number of entangled photons from $n=2$ to $n=3$, while requiring significant experimental efforts, leads to the effective PSF narrowing just by 22\% for the measurement of $G^{(n)}$. The transition from a 3-photon entangled state to a 4-photon one yields only 15\% narrower PSF. Moreover, the actual resolution enhancement is typically smaller than the relative change of the PSF width \cite{pearce_precision_2015,vlasenko2020}, especially for high orders of the correlations, where it may saturate completely. The proposed approach provides a similar magnitude of the resolution increase on the cost of adding a bucket detector to the imaging scheme, which is much simpler than changing the number of entangled photons.

Similar concern about soundness of the results may be elicited by recalling a simple problem of resolving two point sources, commonly investigated theoretically \cite{tsang2016quantum,paur2018tempering,paur2019reading}. For such a simple model situation, the error of inferring the distance $d$ between the sources scales as $\Delta d \propto d^{-1} N^{-1/2}$ \cite{paur2018tempering}, where $N$ is the number of detected events. Therefore, to resolve twice smaller separation of the two sources with the same error, one just needs to perform a 4 times longer experiment and collect $4N$ events. The situation becomes completely different when a more practical multiparametric problem is considered \cite{mikhalychev_efficiently_2019}: the achievable resolution becomes practically insensitive to the data acquisition time (as soon as the number of detected events becomes sufficiently large). For example, for the situation described by the solid black line in Fig.~\ref{fig:Fisher}a, a 100-fold increase of the acquisition time leads to 14\% larger resolution.

\section{Discussions}

We have demonstrated how to enhance resolution of imaging with an $n$-photon entangled state by loosing a photon and measuring the $(n-1)$th-order correlation function instead of the $n$th-order coincidence signal. The resolution gain occurs despite breaking of entanglement as a consequence of the photon loss. We have explained the effect in terms of the effective modification of the remaining photons state when one of the entangled photons is lost. Measurement of $(n-1)$-photon coincidences for an $n$-photon entangled state not only discards some information carried by the ignored $n$th photon, but also makes the resulting signal more informative in the considered imaging experiment. The latter effect prevails for $n > 2$ and leads to increase of the information per measurement and to decrease of the lower bounds for the object inference errors.

The $(n-1)$-photon detection represents a mixture of different possible outcomes for the discarded $n$th photon, including its successful detection at the image plane (resulting in the $n$-photon coincidence signal). The information per a single $(n-1)$-photon detection event is averaged over the discussed possibilities and, for $n > 2$, is larger than the information for a single $n$-photon coincidence event. It means that  certain outcomes for the $n$th photon  provide more information per event than the average value, achieved for $G^{(n-1)}$. We prove that proposition constructively by proposing a hybrid measurement scheme, which provides resolution increase relatively to detection of $(n-1)$-photon coincidences. Intentional detection of a photon outside the optical system, used for imaging of the object, introduces an additional phase shift and increases the sensitivity of the measurement performed with the remaining photons. Our simulations show that the effect can be observed even for $n = 2$, thus making its practical implementation much more realistic. We believe that our observation will pave a way for practical exploitation of entangled states by devising a superresolving imaging  scheme conditioned on detecting photons not only successfully passing through the imaging system, but also those missing it. 

\section{Methods}

\subsection{Expressions for field correlation functions}

For the imaging setup, shown in Fig. \ref{fig:scheme}, the positive-frequency field operators $E(\vec r)$ at the detection plane are connected to the operators $E_0(\vec s)$ of the field illuminating the object as
\begin{equation}
\label{eq:Eplus}
E^{(+)}(\vec r) = \int d^2 \vec s E_0^{(+)}(\vec s) A(\vec s) h(\vec s, \vec r).
\end{equation}

The $n$th-order correlation function for the $n$-photon entangled state (\ref{eq:Psi_n}) is calculated according to the following standard definition:
\begin{equation}
\label{eq:Gn_def}
G^{(n)}(\vec r) = \langle \Psi_n | \left[ E^{(-)} (\vec r)  \right]^n  \left[ E^{(+)} (\vec r)  \right]^n  |\Psi_n \rangle,
\end{equation}
where $E^{(-)}(\vec r) = [E^{(+)}(\vec r)]^+ $ is the negative-frequency field operator. By substitution of Eq. (\ref{eq:Eplus}) into Eq. (\ref{eq:Gn_def}), one can obtain the expression (\ref{eq:Gn}) in the Results section.

The $(n-1)$th-order correlation function is calculated according to the expression
\begin{equation}
\label{eq:Gn-1_def}
G^{(n-1)}(\vec r) = \langle \Psi_n | \left[ E^{(-)} (\vec r)  \right]^{n-1}  \left[ E^{(+)} (\vec r)  \right]^{n-1}  |\Psi_n \rangle ,
\end{equation}
which yields Eq. (\ref{eq:Gn-1}) after substitution of Eq. (\ref{eq:Eplus}).

\subsection{Effective $(n-1)$-photon state}

The density operator, describing the effective $(n-1)$-photon state averaged over the possible ``fates'' of the $n$th photon, discussed in the main text, is
\begin{equation}
\label{eq:rho_n-1}
\rho_{n-1} = \rho_{n-1}^{(1)} +\rho_{n-1}^{(2)} +\rho_{n-1}^{(3)} ,
\end{equation}
where the operators $\rho_{n-1}^{(k)} $ are indexed according to the introduced possibilities and normalized in such a way that $\operatorname{Tr}\rho_{n-1}^{(k)} $ is the probability of the $k$th ``fate''.

According to the approach, discussed in Ref. \cite{bhatti2018generation}, detection of the $n$th photon at the position $\vec r'$ of the detector effectively modifies the states of the remaining $(n - 1)$ photons in the following way:
\begin{equation}
\label{eq:Psi_n-1(1)_def}
|\Psi_{n-1}^{(1)}(\vec r')\rangle \propto E^{(+)} (\vec r ') |\Psi_n\rangle.
\end{equation}
Substitution of Eqs. (\ref{eq:Psi_n}) and (\ref{eq:Eplus}) yields Eq. (\ref{eq:Psi_n-1(1)}).

If we ignore the information about the position $\vec r'$ of the photon detection, the contribution to the averaged density operator (\ref{eq:rho_n-1}) is
\begin{equation}
\label{eq:rho_n-1(1)}
    \rho_{n-1}^{(1)} = \int d^2 \vec r' |\Psi_{n-1}^{(1)}(\vec r')\rangle \langle \Psi_{n-1}^{(1)}(\vec r')|.
\end{equation}

For the possibility, described by Eq. (\ref{eq:Psi_n-1(2)}) and corresponding to the $n$th photon passage outside the aperture of the imaging system, the contribution to the averaged density operator (\ref{eq:rho_n-1}) is
\begin{equation}
\label{eq:rho_n-1(2)}
    \rho_{n-1}^{(2)} = \int_{|\vec k| > k_{max}} d^2 \vec k |\Psi_{n-1}^{(2)}(\vec k)\rangle \langle \Psi_{n-1}^{(2)}(\vec k)|,
\end{equation}
where $k_{max}$ is the maximal transverse momentum transferred by the optical system: $k_{max} = k R / s_o$; $k$ is the wavenumber of the light, $R$ is the radius of the aperture, and $s_o$ is the distance between the object and the lens used for imaging.

Calculating integrals in Eqs. (\ref{eq:rho_n-1(1)}), (\ref{eq:rho_n-1(2)}), and (\ref{eq:rho_n-1(3)}) and taking into account the connection between the PSF shape and $k_{max}$ (see e.g. Ref. \cite{shih_2018_introduction}), one can obtain Eq. (\ref{eq:rho_n-1_final}) for the effective $(n-1)$-photon state.

\subsection{Model of point-spread function}

For the simulations, illustrated by Figs. \ref{fig:simulations}, \ref{fig:Fisher}, and \ref{fig:rate}, we assume for simplicity that the magnification of the optical system is equal to 1, neglect the phase factor in PSF, and use the expression
\begin{multline}
% Sign of "i" in the exponent here and below to be checked!
\label{eq:h}
h(\vec s, \vec r) = \int _{|\vec k| \le k_{max}} d^2 \vec k\, e^{i \vec k \cdot (\vec s + \vec r)} \\{} = 2 \pi k_{max}^2 \operatorname{somb} \left(k_{max} \left|\vec s + \vec r\right|\right),
\end{multline}
where $\operatorname{somb}(x) = 2 J_1(x) / x $, $J_1(x)$ is the first-order Bessel function, and $k_{max}$ is the maximal transverse momentum transferred by the optical system.

\subsection{Quantification of resolution}

Let us assume that a reasonable number of detected coincidence events $N$ in a quantum imaging experiment is limited by the value $N_{max}$ and the acceptable total reconstruction error (see Eq. (\ref{eq:Delta2})) is $\Delta^2 \le 1$. Therefore, Eq. (\ref{eq:Delta2}) implies the following threshold for the trace of the inverse of FIM:
\begin{equation}
\label{eq:FIM_threshold}
 \operatorname{Tr}F^{-1} \le N \Delta^2 \le N_{max}.
\end{equation}
Therefore, one can define the spatial resolution, achievable under the described experimental conditions, as the minimal feature size $d$, for which the condition (\ref{eq:FIM_threshold}) is satisfied. 

\section{Acknowledgements}

All the authors acknowledge financial support from the King Abdullah University of Science and Technology (grant 4264.01), A. M., I. K., and D.M. also acknowledge support from the EU Flagship on Quantum Technologies, project PhoG (820365).

\section{Author contributions}

The theory was conceived by A.M. and D.M. Numerical calculations were performed by A.M. The project was supervised by D.M., D.L.M., and A.M. All the authors participated in the manuscript preparation, discussions and checks of the results. 

\section{Competing interests}

The authors declare no competing interests.

\bibliography{articles}

\end{document}